\documentclass[journal]{IEEEtran}
\usepackage{amsmath,amssymb,amsfonts,bm}
\usepackage{graphicx}
\usepackage[caption=false,font=footnotesize]{subfig}
\usepackage{cite}
\usepackage[colorlinks=false,pdfborder={0 0 0}]{hyperref}
\newcommand{\iv}{\bm{\iota}}
\newcommand{\jj}{\jmath}
\newcommand{\Zb}{\mathbf{Z}}
\newcommand{\Tm}{\mathbf{T}}
\newcommand{\Bm}{\mathbf{B}}
\newcommand{\Am}{\mathbf{A}}
\newcommand{\gv}{\mathbf{g}}
\newcommand{\thv}{\bm{\theta}}
\newcommand{\yv}{\mathbf{y}}
\newcommand{\nv}{\mathbf{n}}
\newcommand{\Xv}{\mathbf{x}}
\newcommand{\Ident}{\mathbf{I}}
\newcommand{\Cth}{\mathbf{C}_{\thv}}
\newcommand{\tr}{\operatorname{tr}}
\newcommand{\E}{\mathbb{E}}

\begin{document}

\title{Channel Estimation for Stacked Intelligent Metasurfaces with Mutual Coupling}

\author{Chandan~Kumar~Sheemar,~\IEEEmembership{Member,~IEEE}, Giovanni~Iacovelli,~\IEEEmembership{Member,~IEEE}, Symeon~Chatzinotas,~\IEEEmembership{Fellow,~IEEE}
\thanks{C. K. Sheemar, G. Iacovelli, and S. Chatzinotas are with the Interdisciplinary Centre for Security, Reliability and Trust (SnT), University of Luxembourg, Luxembourg (e-mail: \{chandankumar.sheemar, giovanni.iacovelli, symeon.chatzinotas\}@uni.lu).}
}

\maketitle

\begin{abstract}
Channel estimation for stacked intelligent metasurfaces (SIMs) has relied on the cascade model, which ignores mutual coupling and inter-layer reflection. Under a coupling-consistent multiport model, we show that the pilots are linear in the unknown channel segments and derive least-squares and linear minimum-mean-square-error estimators, an identifiability condition and an exact error decomposition for the reactance training schedule, and a gradient-based schedule design. Cascade-based estimation floors above $0$~dB normalized error at practical layer spacings and collapses the sum rate below that of the bare feed, whereas the proposed estimators retain about $80\%$ of the perfect-channel-knowledge sum rate.
\end{abstract}


\section{Introduction}
\IEEEPARstart{S}{tacked} intelligent metasurfaces (SIMs) cascade several programmable transmissive layers in front of a compact feed array, so that precoding and combining are performed by the propagating field itself, in the wave domain, before digitization \cite{An2023SIM,Sheemar2026Survey}. Since the number of meta-atoms per layer far exceeds the number of radio-frequency chains, acquiring channel state information (CSI) is the bottleneck of every SIM-aided design, and a growing body of work addresses it: subspace-based estimators with optimized training phase shifts \cite{Yao2024CE}, minimum mean square error (MMSE) estimation under Rician fading \cite{Papazafeiropoulos2025Rician}, hybrid digital--wave-domain estimation \cite{Nadeem2024Hybrid}, nested-tensor \cite{Ginige2025Nested} and parity-dependent tensor \cite{Almeida2026TenSIM} decompositions, and learning-based refinements surveyed in \cite{Sheemar2026Survey}.

Nearly all of these schemes adopt the \emph{cascade model}: each layer is a diagonal mask of unit-modulus phase shifts, and consecutive layers are connected by a scalar, unilateral Rayleigh--Sommerfeld (RS) propagator \cite{An2023SIM}. A parallel, circuit-theoretic literature has shown that this abstraction is not the physics of a compact stack. Physically consistent scattering- and impedance-parameter models of SIMs \cite{Nerini2024PhysSIM,Abrardo2025Multiport} reveal intra-layer mutual coupling, inter-layer reflections, and a load-dependent input impedance seen by the feed. The model of \cite{Iacovelli2026Stacked} captures all of these in one closed-form impedance kernel sampled at the three-dimensional (3-D) port separations, shows that the cascade is the single-pass limit of one matrix inversion under assumptions that fail in compact stacks, and finds that a precoder designed on the cascade can perform worse than open-circuiting the stack. The only coupling-aware SIM estimator is available in \cite{Abrardo2026NF}, which configures a receive SIM as a single analog projector onto a subspace derived from prior location knowledge, for near-field channel and localization estimation. Training over multiple load states, identifiability without prior knowledge, and the error committed by cascade-based estimators remain open \cite{Sheemar2026Survey}, and are addressed here.

\emph{Contributions:} (i) we formulate uplink SIM channel estimation under a coupling-consistent multiport model and show that, although the loaded stack responds nonlinearly to its reactances, the received pilots are \emph{linear} in the unknown feed-side and stack-side channel segments, with a known, load-dependent regressor; the model is exact in the radiating near field, so no planar-wavefront assumption is needed; (ii) we derive least-squares (LS) and linear MMSE (LMMSE) estimators over a schedule of reactance states, a necessary and sufficient identifiability condition on the schedule, the exact mean square error (MSE) and its feed/stack decomposition, the exact prediction error at untrained load states, the high-SNR floor of the LMMSE estimator below the pilot counting bound, and a gradient-based schedule design; (iii) we derive the effective-channel normalized MSE (NMSE) of every linear estimator in closed form, including the bias floor of cascade-based estimation and of model errors in general; and (iv) we show that this bias is minor below $0.1\lambda$ of layer spacing, where the stack is nearly inert, but reaches about $100\%$ NMSE or more from $\lambda/6$ to $\lambda/2$, a range covering the spacings at which a compact SIM delivers its gains \cite{Iacovelli2026Stacked}; that cascade-based CSI collapses the sum rate of load optimization below that of the bare feed; and that the proposed estimators retain about $80\%$ of the perfect-CSI rate and tolerate realistic fabrication errors, $2$--$3$-bit reactances, and multipath.

\emph{Notation:} boldface lowercase (uppercase) letters are vectors (matrices); $(\cdot)^{\mathsf{T}}$, $(\cdot)^{\mathsf{H}}$, $(\cdot)^{*}$ denote transpose, conjugate transpose, and conjugate; $\mathrm{Re}\{\cdot\}$ and $\mathrm{Im}\{\cdot\}$ act entrywise; $\mathrm{diag}(\cdot)$ maps a vector to a diagonal matrix and a matrix to its diagonal; $\jj$ is the imaginary unit with convention $e^{-\jj 2\pi f t}$; $j_\ell,y_\ell,h_\ell=j_\ell+\jj y_\ell$ are spherical Bessel and Hankel functions; $\|\cdot\|_2$ and $\|\cdot\|_{\mathrm{F}}$ are the spectral and Frobenius norms $\sigma_{\min}(\cdot)$ the smallest singular value, $(\cdot)^{+}$ the pseudoinverse, and $\mu_i(\cdot)$ the $i$-th largest eigenvalue; $\mathbf{e}_n$ is the $n$-th canonical vector; $\lambda$ is the wavelength and $\kappa=2\pi/\lambda$.

\section{Coupling-Consistent System Model}\label{sec:model}
We adopt the multiport model of \cite{Iacovelli2026Stacked}, specialized to fixed port positions. A base station (BS) hosts $N_0$ driven ports on the plane $z=0$ and an $L$-layer SIM whose $l$-th sheet carries $N_l$ identical $x$-polarized sub-wavelength elements on the plane $z=z_l$, $0<z_1<\dots<z_L$, with $N_{\mathrm{s}}=\sum_{l\geq1}N_l$ stack ports in total. The mutual impedance between any two ports, on the same sheet or across sheets, is the free-space dyadic Green function projected on the polarization and sampled at their 3-D separation $\bm{\tau}$; in the point-dipole limit and normalized by the radiation resistance $R_r$, it reads
\begin{equation}\label{eq:kernel}
\vartheta(\rho,c)=\tfrac{3}{2}\big[h_0(\rho)-\tfrac{h_1(\rho)}{\rho}+h_2(\rho)\,c^2\big],
\end{equation}
with $\rho=\kappa\|\bm{\tau}\|$ and $c$ the cosine between the polarization and $\bm{\tau}$. Finite element profiles smooth \eqref{eq:kernel} at gaps below ${\sim}0.1\lambda$ \cite{Iacovelli2026Stacked}; nothing below depends on its form beyond the structure of the multiport impedance matrix
\begin{equation}\label{eq:Z}
\Zb = (1+\varepsilon+\jj\zeta_A)\,\Ident_{N} + \bm{\Theta},\qquad N=N_0+N_{\mathrm{s}},
\end{equation}
where $\varepsilon$ and $\zeta_A$ are the normalized ohmic self-resistance and self-reactance and $\bm{\Theta}$ collects \eqref{eq:kernel} off the diagonal. With port currents $\iv=[\iv_0^{\mathsf{T}},\iv_{\mathrm{s}}^{\mathsf{T}}]^{\mathsf{T}}$ and voltages $\mathbf{v}=\Zb\iv$ partitioned into feed (subscript $0$) and stack (subscript $\mathrm{s}$) blocks, and the stack block into sheet blocks $\Zb_{ll'}$, reciprocity gives $\Zb_{\mathrm{s}0}=\Zb_{0\mathrm{s}}^{\mathsf{T}}$ and $\Zb_{ll'}=\Zb_{l'l}^{\mathsf{T}}$: the backward transfer between sheets equals the forward one, which is precisely what a unilateral propagator discards.

Each stack port $n$ is terminated on a tunable reactance $x_n$, i.e., on the load network $\mathbf{M}=\jj\,\mathrm{diag}(\Xv)$, $\Xv=[x_1,\dots,x_{N_{\mathrm{s}}}]^{\mathsf{T}}\in\mathbb{R}^{N_{\mathrm{s}}}$. Substituting the terminal law $\mathbf{v}_{\mathrm{s}}=-\mathbf{M}\iv_{\mathrm{s}}$ into the multiport relation $\mathbf{v}_{\mathrm{s}}=\Zb_{\mathrm{s}0}\iv_0+\Zb_{\mathrm{ss}}\iv_{\mathrm{s}}$ gives the induced currents
\begin{equation}\label{eq:B}
\iv_{\mathrm{s}} = -\Bm(\Xv)\,\iv_0,\quad
\Bm(\Xv)\triangleq\big(\Zb_{\mathrm{ss}}+\jj\,\mathrm{diag}(\Xv)\big)^{-1}\Zb_{\mathrm{s}0},
\end{equation}
which contain every multiple reflection, intra-layer coupling, and feedback term. Since $\mathrm{Re}\{\Zb_{\mathrm{ss}}\}\succeq\varepsilon\Ident$ by passivity, \eqref{eq:B} is well posed for every $\Xv$ when $\varepsilon>0$.

A single-antenna user $k$, modeled as an $x$-polarized dipole at $\mathbf{r}_k$, is one more port of the same multiport: its couplings to the feed and stack ports are the kernel samples $[\gv_{k,0}]_n=\vartheta(\kappa\|\mathbf{r}_k-\mathbf{r}_{0,n}\|,c)$ and $[\gv_{k,\mathrm{s}}]_n=\vartheta(\kappa\|\mathbf{r}_k-\mathbf{r}_{\mathrm{s},n}\|,c)$, with $\mathbf{r}_{0,n}$, $\mathbf{r}_{\mathrm{s},n}$ the port positions and $c$ the polarization cosine of each pair, as in \eqref{eq:kernel}. This is a near-field line-of-sight channel; with scatterers, $\gv_{k,0}$ and $\gv_{k,\mathrm{s}}$ become the user's field at the ports and nothing below changes. As the user's open-circuit voltage is $\gv_{k,0}^{\mathsf{T}}\iv_0+\gv_{k,\mathrm{s}}^{\mathsf{T}}\iv_{\mathrm{s}}$, \eqref{eq:B} yields the effective end-to-end channel, i.e., the user--feed entry of the Schur complement of the loaded multiport with the stack ports eliminated,
\begin{equation}\label{eq:eff}
\tilde{\gv}_k^{\mathsf{T}}(\Xv) = \gv_{k,0}^{\mathsf{T}} - \gv_{k,\mathrm{s}}^{\mathsf{T}}\Bm(\Xv),
\end{equation}
which, by reciprocity, is also the uplink channel from user $k$ to the open-circuit port voltages of the BS. Since $\Zb_{\mathrm{ss}}$ and $\Zb_{\mathrm{s}0}$ depend only on the known geometry, $\Bm(\Xv)$ is computable at the BS for any load state, and \eqref{eq:eff} is linear in the only unknowns, $\gv_{k,0}$ and $\gv_{k,\mathrm{s}}$, with a regressor that the BS steers through $\Xv$.

\emph{Remark 1:} Once $\gv_{k,0}$ and $\gv_{k,\mathrm{s}}$ are estimated, \eqref{eq:eff} predicts the effective channel at \emph{every} load state, which is what a load-optimization loop needs (Corollary~1, Sec.~\ref{sec:rate}), whereas end-to-end estimates \cite{Nadeem2024Hybrid} hold only at the training state. The price of physical consistency is the size of the unknown: every sheet radiates toward the user, so the channel is needed at all $N_0+N_{\mathrm{s}}$ ports, rather than at the outer layer only as in a cascade with known propagators.

\subsection{Training Protocol}\label{sec:train}
Users transmit mutually orthogonal pilots over $T$ slots of a coherence block; after pilot matching, the per-user observation in slot $t$, during which the loads are set to the schedule entry $\Xv_t$, is
\begin{equation}\label{eq:slot}
\yv_t = \tilde{\gv}_k(\Xv_t)\sqrt{p} + \nv_t
= \big(\gv_{k,0}-\Bm_t^{\mathsf{T}}\gv_{k,\mathrm{s}}\big)\sqrt{p} + \nv_t,
\end{equation}
with $\Bm_t\triangleq\Bm(\Xv_t)$, pilot power $p$ (set to $1$ henceforth), and $\nv_t\sim\mathcal{CN}(\mathbf{0},\sigma^2\Ident_{N_0})$. Stacking the slots and defining $\thv\triangleq[\gv_{k,0}^{\mathsf{T}},\gv_{k,\mathrm{s}}^{\mathsf{T}}]^{\mathsf{T}}\in\mathbb{C}^{N}$,
\begin{equation}\label{eq:lin}
\yv = \Am\,\thv+\nv,\qquad
\Am = \begin{bmatrix}\Ident_{N_0} & -\Bm_1^{\mathsf{T}}\\[-1pt] \vdots & \vdots \\[-1pt] \Ident_{N_0} & -\Bm_T^{\mathsf{T}}\end{bmatrix}\in\mathbb{C}^{TN_0\times N}.
\end{equation}
The training states are reactance vectors, not phase masks: the hardware knob is $\Xv$, to which the coupled stack responds only through $\Bm(\Xv)$. For intuition, consider an isolated atom (coupling neglected): one port of normalized self-impedance $a+\jj\zeta_A$, $a\triangleq1+\varepsilon$, terminated on one lossless reactance $x_n$. Its induced current per unit incident voltage, $\psi(x_n)\triangleq-1/(a+\jj(\zeta_A+x_n))$, traces a circle through the origin as $x_n$ sweeps $\mathbb{R}$, locking amplitude and phase, so no lossless one-port load realizes the unit-modulus phase shifts assumed in \cite{Yao2024CE,Papazafeiropoulos2025Rician,Nadeem2024Hybrid}. We draw training reactances as
\begin{equation}\label{eq:locus}
x_{t,n} = -\zeta_A + a\tan u_{t,n},\qquad u_{t,n}\sim\mathcal{U}(-u_{\max},u_{\max}).
\end{equation}
Under \eqref{eq:locus} the isolated response is $-(\cos u_{t,n}/a)\,e^{-\jj u_{t,n}}$, whose phase is thus sampled uniformly about the element resonance $u=0$, whereas uniformly drawn reactances would crowd near the open circuit, where atoms are nearly inert. In the coupled stack, \eqref{eq:locus} is merely a convenient prior over $\Xv$: by Proposition~1 below, it yields identifiability almost surely, as would any absolutely continuous distribution, and Sec.~\ref{sec:design} optimizes the schedule.

\section{Estimators, Identifiability, and Schedule Design}\label{sec:est}
\subsection{Least Squares and Identifiability}
When $\Am$ has full column rank, the LS estimate and its error covariance are
\begin{equation}\label{eq:ls}
\hat{\thv}_{\mathrm{LS}} = (\Am^{\mathsf{H}}\Am)^{-1}\Am^{\mathsf{H}}\yv,\quad
\mathrm{MSE} = \sigma^2\,\tr\big[(\Am^{\mathsf{H}}\Am)^{-1}\big],
\end{equation}
which is unbiased and attains the Cram\'er--Rao bound (CRB) for deterministic $\thv$ \cite{Kay1993}.

\emph{Proposition 1 (Identifiability):} $\Am$ in \eqref{eq:lin} has full column rank $N_0+N_{\mathrm{s}}$ if and only if
\begin{equation}\label{eq:ident}
\operatorname{rank}\big[\,\Bm_2-\Bm_1,\;\Bm_3-\Bm_1,\;\dots,\;\Bm_T-\Bm_1\,\big] = N_{\mathrm{s}},
\end{equation}
so that $(T-1)N_0\geq N_{\mathrm{s}}$, i.e., $T\geq 1+\lceil N_{\mathrm{s}}/N_0\rceil$, is necessary. Moreover, if one schedule of length $T$ satisfies \eqref{eq:ident}, then almost every schedule drawn from \eqref{eq:locus}, or from any absolutely continuous distribution, does.

\emph{Proof:} $\Am\thv=\mathbf{0}$ iff $\gv_{k,0}=\Bm_t^{\mathsf{T}}\gv_{k,\mathrm{s}}$ for all $t$, i.e., iff $(\Bm_t-\Bm_1)^{\mathsf{T}}\gv_{k,\mathrm{s}}=\mathbf{0}$ for all $t$ and $\gv_{k,0}=\Bm_1^{\mathsf{T}}\gv_{k,\mathrm{s}}$; the null space is trivial iff the stacked differences have rank $N_{\mathrm{s}}$, which bounds $T$ by counting. For genericity, every entry of $\Bm(\Xv)$ is a rational function of $\Xv$ by \eqref{eq:B} and Cramer's rule, so the minors of the matrix in \eqref{eq:ident} are rational in $(\Xv_1,\dots,\Xv_T)$, and the rank condition fails only on the common zero set of all top-order minors; these are real-analytic and, once one full-rank schedule exists, not all identically zero, so that set has Lebesgue measure zero. In the setting of Sec.~\ref{sec:num}, such a schedule exists already at the counting bound $T=14$, for every spacing considered. $\blacksquare$

\emph{Proposition 2 (MSE decomposition):} Let $\bar{\Bm}=\frac{1}{T}\sum_t\Bm_t$, $\bm{\Delta}_t=\Bm_t-\bar{\Bm}$, and $\mathbf{S}=\sum_t\bm{\Delta}_t^{*}\bm{\Delta}_t^{\mathsf{T}}$. Under \eqref{eq:ident}, the LS errors of the stack and feed segments are
\begin{equation}\label{eq:decomp}
\mathrm{MSE}_{\mathrm{s}}=\sigma^2\tr(\mathbf{S}^{-1}),\quad
\mathrm{MSE}_{0}=\sigma^2\big[\tfrac{N_0}{T}+\tr(\bar{\Bm}^{\mathsf{T}}\mathbf{S}^{-1}\bar{\Bm}^{*})\big].
\end{equation}

\emph{Proof:} $\Am^{\mathsf{H}}\Am=\big[\begin{smallmatrix}T\Ident & -T\bar{\Bm}^{\mathsf{T}}\\ -T\bar{\Bm}^{*} & \sum_t\Bm_t^{*}\Bm_t^{\mathsf{T}}\end{smallmatrix}\big]$; its leading block has Schur complement $\mathbf{S}$, and block inversion gives
\begin{equation}\label{eq:blockinv}
(\Am^{\mathsf{H}}\Am)^{-1}=\begin{bmatrix}T^{-1}\Ident+\bar{\Bm}^{\mathsf{T}}\mathbf{S}^{-1}\bar{\Bm}^{*} & \bar{\Bm}^{\mathsf{T}}\mathbf{S}^{-1}\\ \mathbf{S}^{-1}\bar{\Bm}^{*} & \mathbf{S}^{-1}\end{bmatrix},
\end{equation}
whose diagonal blocks give \eqref{eq:decomp}. $\blacksquare$

The stack segment is thus learned only from the \emph{variation} of the regressor across load states, $\mathbf{S}\succ\mathbf{0}$ being exactly \eqref{eq:ident}; in the setting of Sec.~\ref{sec:num} at inter-layer spacing $d=\lambda/4$, it carries about $99\%$ of the MSE, six to seven times the feed error per entry.

\emph{Corollary 1 (Prediction at any load state):} The exact-model LS prediction $\mathbf{C}(\Xv)\hat{\thv}_{\mathrm{LS}}$ of the effective channel \eqref{eq:eff}, with $\mathbf{C}(\Xv)=[\Ident_{N_0},-\Bm^{\mathsf{T}}(\Xv)]$, is unbiased at every state $\Xv$, trained or not, with MSE
\begin{equation}\label{eq:pred}
\sigma^2\Big[\frac{N_0}{T}+\tr\big((\Bm(\Xv)-\bar{\Bm})^{\mathsf{T}}\mathbf{S}^{-1}(\Bm(\Xv)-\bar{\Bm})^{*}\big)\Big].
\end{equation}

\emph{Proof:} The error covariance is $\sigma^2\mathbf{C}(\Xv)(\Am^{\mathsf{H}}\Am)^{-1}\mathbf{C}^{\mathsf{H}}(\Xv)$; inserting \eqref{eq:blockinv} and collecting terms gives \eqref{eq:pred}. $\blacksquare$

The prediction error grows with the $\mathbf{S}^{-1}$-weighted distance of $\Bm(\Xv)$ from the training mean, and $\Bm=\mathbf{0}$ recovers $\mathrm{MSE}_0$: a schedule should span the load region in which the optimizer of Sec.~\ref{sec:rate} operates. This makes Remark~1 quantitative.

With $K$ users and orthogonal pilots, all users are estimated in the same $T$ load states at $K$-fold pilot length, and the per-user problems decouple. Online, both estimators cost one $N\times TN_0$ matrix--vector product per user and coherence block, their matrices being precomputed per geometry and schedule (and noise level, for LMMSE) from $T$ solves of size $N_{\mathrm{s}}$ and one inversion of size $N$ (LS) or $TN_0$ (LMMSE). Since \eqref{eq:ls} attains the CRB among unbiased estimators, the high-SNR floors of the mismatched schemes in Sec.~\ref{sec:num} are due to bias, not to estimator suboptimality; at low SNR, a biased scheme may fall below the bound.

\subsection{LMMSE with Statistical CSI}
Users confined to a known coverage sector induce the prior correlation $\Cth=\E[\thv\thv^{\mathsf{H}}]$, computable offline from \eqref{eq:kernel} by Monte Carlo over the sector. The LMMSE estimate is
\begin{equation}\label{eq:lmmse}
\hat{\thv}_{\mathrm{LMMSE}} = \Cth\Am^{\mathsf{H}}\big(\Am\Cth\Am^{\mathsf{H}}+\sigma^2\Ident_{TN_0}\big)^{-1}\yv .
\end{equation}
\emph{Proposition 3 (LMMSE below the counting bound):} Let $\mathbf{K}=\Am\Cth^{1/2}$. The LMMSE error covariance $\mathbf{E}(\sigma^2)=\Cth-\Cth\Am^{\mathsf{H}}(\Am\Cth\Am^{\mathsf{H}}+\sigma^2\Ident)^{-1}\Am\Cth$ decreases, in the Loewner order, as $\sigma^2\downarrow0$ to
\begin{equation}\label{eq:floor}
\mathbf{E}_\infty=\Cth^{1/2}(\Ident_N-\mathbf{K}^{+}\mathbf{K})\Cth^{1/2},
\end{equation}
which vanishes iff $\Am$ is injective on the range of $\Cth$, so that $TN_0\geq\operatorname{rank}\Cth$ is necessary; moreover, $\tr\mathbf{E}_\infty\geq\sum_{i>TN_0}\mu_i(\Cth)$.

\emph{Proof:} $\mathbf{E}(\sigma^2)=\Cth^{1/2}[\Ident-\mathbf{K}^{\mathsf{H}}(\mathbf{K}\mathbf{K}^{\mathsf{H}}+\sigma^2\Ident)^{-1}\mathbf{K}]\Cth^{1/2}$, and the bracket decreases to $\Ident-\mathbf{K}^{+}\mathbf{K}$, the projector onto the null space of $\mathbf{K}$, which annihilates $\Cth^{1/2}$ iff $\operatorname{rank}\mathbf{K}=\operatorname{rank}\Cth$. Since $\tr\mathbf{E}_\infty=\tr\Cth-\tr(\mathbf{K}^{+}\mathbf{K}\Cth)$ and $\mathbf{K}^{+}\mathbf{K}$ is a projector of rank at most $TN_0$, Ky Fan's maximum principle gives the bound. $\blacksquare$

The overhead is thus governed by the effective rank of the prior, not by $N_{\mathrm{s}}$: in our setting $99\%$ of the prior energy lies in $9$ of $N=54$ eigendirections, and the normalized floor $\tr\mathbf{E}_\infty/\tr\Cth$ is $-9.8$, $-25.7$, and $-54.2$~dB for $T=2$, $4$, and $8$, against $T=14$ for LS.

\subsection{Gradient-Based Schedule Design}\label{sec:design}
The MSE in \eqref{eq:ls} depends on the schedule through $\Bm_t$. Differentiating $\Tm(\Zb_{\mathrm{ss}}+\jj\,\mathrm{diag}(\Xv))=\Ident$ gives $\partial\Tm/\partial x_n=-\jj\Tm\mathbf{e}_n\mathbf{e}_n^{\mathsf{T}}\Tm$, hence
\begin{equation}\label{eq:grad}
\frac{\partial\Bm}{\partial x_n} = -\jj\,\Tm\,\mathbf{e}_n\mathbf{e}_n^{\mathsf{T}}\Bm,\qquad
\Tm=\big(\Zb_{\mathrm{ss}}+\jj\,\mathrm{diag}(\Xv)\big)^{-1}.
\end{equation}
Let $\mathbf{G}=\Am^{\mathsf{H}}\Am$ and $f=\tr(\mathbf{G}^{-1})$. Since $\mathrm{d}f=-2\,\mathrm{Re}\,\tr(\mathbf{G}^{-2}\Am^{\mathsf{H}}\mathrm{d}\Am)$, \eqref{eq:grad} gives the gradient with respect to the reactances of state $t$ in closed form,
\begin{equation}\label{eq:gradf}
\nabla_{\Xv_t} f = 2\,\mathrm{Im}\big\{\mathrm{diag}\big(\Tm_t\mathbf{V}_t\Bm_t^{\mathsf{T}}\big)\big\},
\end{equation}
where $\Tm_t=\Tm(\Xv_t)$, $\mathbf{V}_t$ is the block of $\mathbf{G}^{-2}\Am^{\mathsf{H}}$ with stack rows and slot-$t$ columns, and the symmetry $\Tm_t=\Tm_t^{\mathsf{T}}$, inherited from reciprocity, has been used. All $N_{\mathrm{s}}$ partials of a state thus cost one $N_{\mathrm{s}}\times N_{\mathrm{s}}$ inversion. We minimize $f$ in the variables $u_{t,n}$ of \eqref{eq:locus}, box-constrained to $[-u_{\max},u_{\max}]$, by projected gradient descent with Armijo backtracking from a random draw of \eqref{eq:locus}. The problem is non-convex, $\Bm$ being rational in $\Xv$, but independent random initializations converge to A-optimality values within $0.2$~dB of each other, in about $6$~s for $T=16$, $N_{\mathrm{s}}=50$. By Proposition~2, what matters is how evenly the regressor variation $\mathbf{S}$ covers all $N_{\mathrm{s}}$ directions, not its size: at $d=\lambda/4$ the optimizer lowers the condition number of $\mathbf{S}$ from $1.2\cdot10^{4}$ to $50$, while $\tr\mathbf{S}$ even drops from $28.3$ to $25.3$. The design is performed offline once per geometry, and every optimized state is realizable by construction.

\section{The Cost of the Cascade Model}\label{sec:mismatch}
Suppose the estimator relies on the cascade model, as in prior SIM estimation work. In the multiport language, the cascade keeps in \eqref{eq:B} only the forward nearest-neighbor blocks: sheet $l$ is excited by sheet $l-1$ alone through $\Zb_{l,l-1}$ (sheet $1$ by the feed through $\Zb_{10}$), while the backward blocks $\Zb_{l,l+1}$ and the direct blocks $\Zb_{l0}$, $l\geq2$, are discarded. The currents $\iv_l$ of sheet $l$ then obey $\iv_l=\bm{\Psi}_l\Zb_{l,l-1}\iv_{l-1}$, $\bm{\Psi}_l=-(\Zb_{ll}+\jj\,\mathrm{diag}(\Xv^{(l)}))^{-1}$, with $\Xv^{(l)}$ the reactances of sheet $l$, yielding a mismatched regressor $\breve{\Bm}_t$ and matrix $\breve{\Am}$. The \emph{phase-mask cascade} additionally neglects intra-layer coupling and replaces $\bm{\Psi}_l$ by $\mathrm{diag}(e^{\jj\arg\psi(x^{(l)}_n)})$, with $\psi$ the isolated response of Sec.~\ref{sec:train}: this is the model under which the training designs of \cite{Yao2024CE,Papazafeiropoulos2025Rician,Ginige2025Nested,Almeida2026TenSIM} optimize. Both variants keep the exact kernel blocks as propagators and, unlike the cascade of Remark~1, estimate the same $\thv\in\mathbb{C}^N$, so they are credited with the correct polarized transfer and with the user's coupling to every port: their errors are purely structural. The mismatched LS estimate under the true observations \eqref{eq:lin} is
\begin{equation}\label{eq:bias}
\breve{\thv} = (\breve{\Am}^{\mathsf{H}}\breve{\Am})^{-1}\breve{\Am}^{\mathsf{H}}(\Am\thv+\nv),
\end{equation}
with bias $\bm{\beta}=\breve{\Am}^{+}(\Am-\breve{\Am})\thv$, $\breve{\Am}^{+}\triangleq(\breve{\Am}^{\mathsf{H}}\breve{\Am})^{-1}\breve{\Am}^{\mathsf{H}}$, independent of the noise: both $\breve{\thv}$ and the effective channel it reconstructs through its own $\breve{\Bm}(\Xv)$ carry an NMSE floor at every SNR. The same expression governs any model error, e.g., fabrication tolerances or misestimated $\varepsilon$ and $\zeta_A$, and, for full-column-rank $\breve{\Am}$, bounds the parameter floor as
\begin{equation}\label{eq:biasbound}
\|\bm{\beta}\|/\|\thv\|\leq\|\Am-\breve{\Am}\|_2\,/\,\sigma_{\min}(\breve{\Am}),
\end{equation}
which favors well-conditioned schedules such as those of Sec.~\ref{sec:design}; what sets the cascade apart from such errors is the size of $\Am-\breve{\Am}$. Both discarded families contribute to it: in the setting of Sec.~\ref{sec:num} at $d=\lambda/4$, restoring only the direct block $\Zb_{20}$ lowers the noiseless floor from $+2.7$ to $-2.4$~dB, and restoring only the backward block $\Zb_{12}$ to $+0.7$~dB, so neither truncation alone is benign.

\emph{Effective-channel NMSE in closed form:} Every estimator above is linear, $\hat{\thv}=\mathbf{W}\yv$, and predicts the effective channel at a load state $\Xv_v$ through its own map $\hat{\mathbf{C}}_v=[\Ident_{N_0},-\hat{\Bm}^{\mathsf{T}}(\Xv_v)]$, whereas the truth is $\mathbf{C}_v\thv$ with $\mathbf{C}_v=\mathbf{C}(\Xv_v)$. Averaging over the noise and over users with correlation $\Cth$, the NMSE over a set of states $\{\Xv_v\}$ is
\begin{equation}\label{eq:nmse}
\mathrm{NMSE}=\frac{\sum_v\big[\tr(\mathbf{D}_v\Cth\mathbf{D}_v^{\mathsf{H}})+\sigma^2\|\hat{\mathbf{C}}_v\mathbf{W}\|_{\mathrm{F}}^2\big]}{\sum_v\tr(\mathbf{C}_v\Cth\mathbf{C}_v^{\mathsf{H}})},
\end{equation}
with $\mathbf{D}_v=\hat{\mathbf{C}}_v\mathbf{W}\Am-\mathbf{C}_v$. For the exact LS, $\mathbf{W}=\Am^{+}$ and $\mathbf{D}_v=\mathbf{0}$, so \eqref{eq:nmse} decays as $\sigma^2$, consistently with Corollary~1; for the LMMSE, $\mathbf{W}=\Cth\Am^{\mathsf{H}}(\Am\Cth\Am^{\mathsf{H}}+\sigma^2\Ident)^{-1}$; for the cascades, $\mathbf{W}=\breve{\Am}^{+}$ and $\hat{\Bm}=\breve{\Bm}$, and the first term is the noise-independent floor of the effective channel itself.

\section{Numerical Results}\label{sec:num}
\begin{figure}[t]
\centering
\includegraphics[width=\columnwidth]{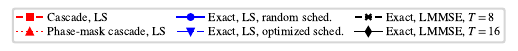}\\[-2pt]
\subfloat[Versus SNR, $d=\lambda/4$]{\includegraphics[width=0.49\columnwidth]{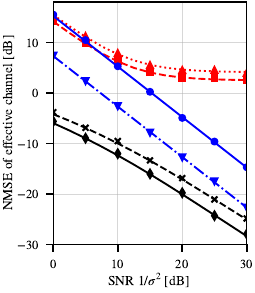}\label{fig:snr}}\hfill
\subfloat[Versus spacing $d$, $30$~dB SNR]{\includegraphics[width=0.49\columnwidth]{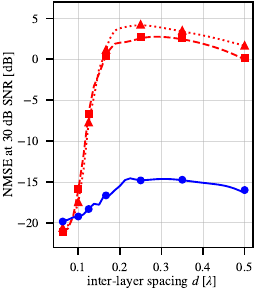}\label{fig:h}}
\caption{Effective-channel NMSE; lines: analysis \eqref{eq:nmse}, markers: Monte Carlo. ``Exact'' denotes estimators built on the coupling-consistent model \eqref{eq:B}--\eqref{eq:lin}; $T=16$ unless marked; (b) uses the random schedule.}
\label{fig:one}
\end{figure}

We follow the scenario of \cite{Iacovelli2026Stacked}: $N_0=4$ feed ports on a central $\lambda/2$ square at $z=0$; $L=2$ sheets of $5\times5$ atoms on a $\lambda/2$ grid ($N_{\mathrm{s}}=50$), the first at standoff $z_1=\lambda/2$ and the second $d$ beyond it; $\varepsilon=0.05$, $\zeta_A=0.58$. Users are dropped uniformly in a $\pm4\lambda$ lateral box, $8\lambda$--$14\lambda$ beyond the last sheet, within the Fraunhofer distance (${\geq}16\lambda$) of the stack; $\thv$ is normalized to $\|\thv\|^2=N$, and $\mathrm{SNR}=1/\sigma^2$. Training uses $T=16$ (counting bound $T=14$) and $u_{\max}=1.35$; the LMMSE prior is estimated from $3000$ kernel drops. The NMSE is that of the effective channel at $10$ random validation load states: in Fig.~\ref{fig:one}, lines show \eqref{eq:nmse} with $20000$-drop statistics (LMMSE weights from the $3000$-drop prior), markers show Monte Carlo averages over $60$ drops and $4$ (a) or $2$ (b) noise realizations; the two agree within $0.5$~dB.

\subsection{Estimation Accuracy}\label{sec:acc}
In Fig.~\ref{fig:one}(a), the exact-model LS estimator is consistent, reaching $-14.7$~dB at $30$~dB SNR, and the optimized schedule gains $8$~dB at equal overhead. The LMMSE estimator reaches $-27.9$~dB, and $-24.9$~dB with $T=8$, below the counting bound, as Proposition~3 predicts; a wider prior sector containing the true one ($\pm8\lambda$, $5\lambda$--$20\lambda$) still gives $-25.1$~dB (not shown). Both cascades saturate at the floors predicted by \eqref{eq:nmse}, $+2.5$ and $+4.1$~dB: their estimate is farther from the truth than the zero vector. In Fig.~\ref{fig:one}(b), the cascade error is minor below $0.1\lambda$, where the reactive kernel, growing as $\|\bm{\tau}\|^{-3}$, makes a resonant stack choke its own currents \cite{Iacovelli2026Stacked} and leaves the direct link dominant (indicative: finite element profiles soften the kernel there). It crosses $0$~dB at $d\approx0.16\lambda$, peaks at $+2.8$~dB near $0.29\lambda$, and stays near $0$~dB at $\lambda/2$, $12$--$18$~dB above the exact LS, whose error varies by only $5$~dB.

Multipath leaves this picture unchanged: with $Q=10$ point scatterers re-radiating as dipoles (uniform in $\pm6\lambda$ laterally, $3\lambda$--$16\lambda$ beyond the last sheet, random complex gains, scattered energy $10$ times the line-of-sight energy in every drop), the $99\%$-energy rank of the prior grows from $9$ to $19$, and at $30$~dB SNR the optimized-schedule LS and the LMMSE estimator, with the prior recomputed, reach $-23.1$ and $-24.6$~dB, the cascades $+2.2$ and $+4.4$~dB ($40$ users, $2$ noise realizations each).

\subsection{Robustness to Model Errors and Quantized Reactances}\label{sec:robust}
In Fig.~\ref{fig:two}(b), the true geometry is perturbed by i.i.d.\ Gaussian jitter of standard deviation $\sigma_p$ per coordinate on every port, or the true $\zeta_A$ is off by $\pm10\%$, while the estimator keeps the nominal model and optimized schedule ($8$ draws, $15$ users, $2$ noise realizations, paired across cases). As \eqref{eq:bias} shows, such errors matter only once the noise falls below the bias they cause: at $20$~dB SNR the loss is at most $0.7$~dB for $\sigma_p\leq0.005\lambda$ (about $54~\mu$m at $28$~GHz, a typical printed-circuit tolerance) and for the $\zeta_A$ error; at $30$~dB, the nominal $-22.7$~dB becomes $-19.0$, $-14.7$, and $-9.3$~dB for $\sigma_p=0.005\lambda$, $0.01\lambda$, and $0.02\lambda$, still about $12$~dB below the cascade floor. With $b$-bit reactances ($2^b$ levels of $u$ equally spaced on $[-u_{\max},u_{\max}]$), projecting the optimized schedule and running two sweeps of coordinate descent keeps full rank and costs $0.6$, $1.4$, and $10.7$~dB at $30$~dB SNR for $b=3$, $2$, and $1$.

\subsection{Sum Rate with Estimated CSI}\label{sec:rate}
\begin{figure}[t]
\centering
\subfloat[Sum rate with estimated CSI]{\includegraphics[width=0.49\columnwidth]{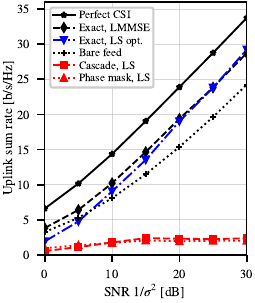}}\hfill
\subfloat[Robustness to model errors]{\includegraphics[width=0.49\columnwidth]{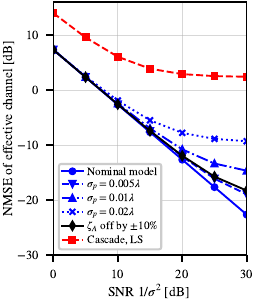}}
\caption{Results at $d=\lambda/4$. (a) Uplink sum rate, $K=3$ users, no training overhead deducted; reactances and MMSE combiners are designed on each scheme's predicted channel and scored on the true one; the bare feed uses perfect CSI. (b) NMSE of the optimized-schedule LS estimator when the true stack deviates from its nominal model.}
\label{fig:two}
\end{figure}

To test Corollary~1 in use, $K=3$ users are estimated as above (data SNR equal to pilot SNR), and each scheme maximizes $F(\Xv)=\log_2\det(\Ident_K+\sigma^{-2}\hat{\mathbf{H}}^{\mathsf{H}}\hat{\mathbf{H}})$ of its own prediction $\hat{\mathbf{H}}(\Xv)=\hat{\mathbf{G}}_0-\Bm^{\mathsf{T}}(\Xv)\hat{\mathbf{G}}_{\mathrm{s}}$ of the effective channels, where $\hat{\mathbf{G}}_0\in\mathbb{C}^{N_0\times K}$ and $\hat{\mathbf{G}}_{\mathrm{s}}\in\mathbb{C}^{N_{\mathrm{s}}\times K}$ collect the estimated segments, by projected gradient ascent over the box of \eqref{eq:locus} from three random starts. Since $\mathrm{d}F=\frac{2}{\sigma^2\ln2}\mathrm{Re}\,\tr(\mathbf{M}\hat{\mathbf{H}}^{\mathsf{H}}\mathrm{d}\hat{\mathbf{H}})$ with $\mathbf{M}=(\Ident_K+\sigma^{-2}\hat{\mathbf{H}}^{\mathsf{H}}\hat{\mathbf{H}})^{-1}$ and $\mathrm{d}\hat{\mathbf{H}}=-\mathrm{d}\Bm^{\mathsf{T}}\hat{\mathbf{G}}_{\mathrm{s}}$, \eqref{eq:grad} gives, for the exact model,
\begin{equation}\label{eq:rategrad}
\nabla_{\Xv}F=-\tfrac{2}{\sigma^2\ln2}\,\mathrm{Im}\big\{\mathrm{diag}\big(\Tm\hat{\mathbf{G}}_{\mathrm{s}}\mathbf{M}\hat{\mathbf{H}}^{\mathsf{H}}\Bm^{\mathsf{T}}\big)\big\},
\end{equation}
and the cascade models are differentiated likewise. The BS then applies linear MMSE combining designed on $\hat{\mathbf{H}}(\Xv)$, and the sum rate is scored on the true channel. Fig.~\ref{fig:two}(a) averages $40$ drops, deducts no training overhead, and grants the bare feed (stack open-circuited, $\Tm=\mathbf{0}$) perfect CSI. At $20$~dB SNR, optimizing the reactances with perfect CSI raises the sum rate from $15.4$ to $23.9$~b/s/Hz; the exact-model LMMSE and optimized-schedule LS estimates retain $19.4$ and $19.0$~b/s/Hz, about $80\%$ of the perfect-CSI rate and half of the stack's gain, whereas cascade-based CSI never exceeds $2.4$~b/s/Hz, $4.5$ to $10$ times below the bare feed. Charging the $KT=48$ pilot slots through the pre-log factor $1-KT/\tau_{\mathrm{c}}$ ($1-K/\tau_{\mathrm{c}}$ for the bare feed), the LMMSE-estimated stack wins once the coherence block exceeds $\tau_{\mathrm{c}}\approx210$--$310$ symbols across the SNR range (e.g., $17.5$ vs.\ $15.3$~b/s/Hz at $20$~dB for $\tau_{\mathrm{c}}=500$), i.e., for low-mobility users; $T=8$ does not lower this threshold.

\section{Conclusion}
This letter formulated channel estimation for SIMs under a mutual-coupling-consistent multiport model, in which the pilots are linear in the unknown channel segments. The LS and LMMSE estimators come with an identifiability condition, an exact error decomposition and prediction error at untrained load states, the LMMSE floor below the counting bound, and a closed-form gradient that optimizes the schedule over realizable reactances. Cascade-based estimation incurs a bias floor, predicted in closed form, of about $100\%$ NMSE or more from $\lambda/6$ to $\lambda/2$ and collapses the sum rate of load optimization below that of the bare feed, whereas the proposed estimators retain about $80\%$ of the perfect-CSI rate, paying off once the coherence block exceeds a few hundred symbols, and tolerate position errors up to $0.005\lambda$, $2$--$3$-bit reactances, and multipath. Since the linear model \eqref{eq:lin}, the estimators \eqref{eq:ls} and \eqref{eq:lmmse}, the rank condition \eqref{eq:ident}, and Proposition~2 require only that $\Zb_{\mathrm{ss}}+\mathbf{M}$ be invertible and symmetric, they hold for any lossless reciprocal beyond-diagonal load network $\mathbf{M}$ \cite{Li2023BDRIS}; redesigning \eqref{eq:locus} and \eqref{eq:grad}--\eqref{eq:gradf} for such networks, and wideband kernels, are left for future work.

\end{document}